# A Survey of Software Reliability Models

Ganesh Pai
Department of ECE
University of Virginia, VA
`g.pai@ieee.org`

Dec. 6, 2002

**Abstract**

Software reliability analysis is performed at various stages during the process of engineering software as an attempt to evaluate if the software reliability requirements have been (or might be) met. In this report, I present a summary of some fundamental black-box and white-box software reliability models. I also present some general shortcomings of these models and suggest avenues for further research.

# 1   Introduction

In the design of complex systems today, an increasing amount of functionality is delegated to software. This is primarily because software is an intellectual product, which is not bounded by the constraints of the physical world in the sense that an equivalent hardware system would be. Since software always operates in the context of a larger system, the dependability requirements on the system filter down to the software component(s) and become the desired software dependability requirements.

Software reliability is one of the important parameters of software quality and system dependability. It is defined as the probability of failure-free software operation in a specified environment for a specified period of time [Lyu96]. A software *failure* occurs when the behavior of the software departs from its specifications, and it is the result of a software *fault*, a design defect, being activated by certain input to the code during its execution.

Software reliability analysis is performed at various stages during the process of engineering software, for a system, as an attempt to evaluate if the software reliability requirements have been (or might be) met. The analysis results not only provide feedback to the designers but also become a measure of software quality. There are two activities related to software reliability analysis: estimation and prediction. In either activity, statistical inference techniques and reliability models are applied to failure data obtained from testing or during operation to measure software reliability. However, estimation is usually retrospective and it is performed to determine achieved reliability from a point in the past to the present time. The prediction activity, on the other hand, parameterizes reliability models used for estimation and utilizes the available data to predict future reliability. In general, software reliability models can be classified as being *black box* models and *white box* models. The difference between the two is simply that the white box models consider the structure of the software in estimating reliability, while the black box models do not.

In this report, I present a summary of some fundamental black box and white box software reliability models. I have not endeavored to survey all the models that exist in the literature simply because there are far too many of them. Furthermore, many of the models are extrapolations of the fundamental models that are presented here. This is especially true of black box models. The remainder of the report is organized as follows:

In section 2, black box models and related terminology are introduced. Five models from this category: the Jelinski-Moranda de-eutrophication model, the Goel-Okumoto non-homogeneous Poisson process (NHPP) model, the Musa basic execution time model, the enhanced NHPP (ENHPP) model [Gokhale96] and the Littlewood-Verrall bayesian model, are described along with their basic assumptions, data requirements and model output.

In section 3, I present white box models and a high level taxonomy of these models. Then I describe two models in this category that appeared in the literature in the recent past: Krishnamurthy and Mathur's path-based model [Mathur97] and Gokhale *et al.*'s state-based model [Gokhale98].

In section 4, I describe a new theory proposed by Bishop *et al.* [Bishop96, Bishop02] that attempts to relate residual fault content in software to a worst case bound on the failure rate of the software faults and inturn to software reliability.

In Section 5, I conclude the report with some observations. In this section, I also attempt to identify general shortcomings of software reliability models and suggest avenues for further, possibly fruitful, research.

# 2 Black box software reliability models

In the software development process, it is very typical to end up with a product that has many design defects, *i.e.* faults, or popularly known as bugs. For a certain input to the software these faults are activated, resulting in a deviation of the software behavior from its specified behavior *i.e.* a failure. Once failures are detected through the testing process and the corresponding fault(s) are located, then assuming that these faults are perfectly fixed, *i.e.* the process of fixing a fault did not introduce a new fault, software reliability increases. If the failure data is recorded either in terms of number of failures observed per given time period or in terms of the time between failures, statistical models can be used to identify the trend in the recorded data, reflecting the growth in reliability. Such models are known as software reliability growth models (SRGMs) or growth models in general. They are used to, both, predict and estimate software reliability.

All SRGMs are of the black box type since they only consider failure data, or metrics that are gathered if testing data are not available. Black box models do not consider the internal structure of the software in reliability estimation and are called as such because they consider software as a monolithic entity, a black box.

In the subsequent portions of this section, five SRGMs are presented. These are namely the Jelinski-Moranda de-eutrophication model, the Goel-Okumoto non-homogeneous Poisson process (NHPP) model, the Musa basic execution time model, the enhanced NHPP (ENHPP) model and the Littlewood-Verrall bayesian model.

## 2.1 Terminology

Some of the terms commonly used in relation to SRGMs are listed in Table 1. Other terms are also used in different models and are explained as they are encountered.

| *Term* | *Explanation* |
|---|---|
| $M(t)$ | The total number of failures experienced by time *t*. |
| $\mu(t)$ | Mean value function for an SRGM. This represents the expectation of the number of failures expected by time t as estimated by the model. Therefore, we have $\mu(t) = E[M(t)]$. |
| $\lambda(t)$ | Failure intensity, representing the derivative of the mean value function. There-fore we have $\lambda(t) = \mu'(t)$. |
| $Z(\Delta t / t_{i-1})$ | Hazard rate of the software, which represents the probability density of experiencing the $i^{th}$ failure at $t_{i-1} + \Delta t$ given that that $(i-1)^{st}$ failure occurred at $t_{i-1}$. |
| $z(t)$ | Per-fault hazard rate, which represents the probability that a fault, that had not been activated so far, will cause a failure instantaneously when activated. This term is usually assumed to be a constant ($\varphi$) by many of the models. |
| $N$ | Initial number of faults present in the software prior to testing. |

Table 1: Terminology common to SRGMs

Data that are generally supplied to SRGMs are either times between failures $\{\Delta t_1, \Delta t_2, \Delta t_3, ...\}$ or the times at which failure occurred $\{t_1, t_2, t_3, ...\}$. All the models presented here make a common assumption of independence between failures. However, there are frameworks [Trivedi00] that can incorporate statistical dependence between failures. These frameworks are not considered in this survey.

## 2.2 The Jelinski-Moranda Model

Developed in 1972, the Jelinski-Mornada de-eutrophication model (J-M) is one of the first software reliability models.

The model assumes that:
- $N$ initial faults in the code prior to testing are a fixed but known value.
- Failures are not correlated and the times between failures are independent and exponentially distributed random variables.
- Fault removal on failure occurrences is instantaneous and does not introduce any new faults into the software under test
- The hazard rate $z(t)$ of each fault is time invariant and a constant ($\varphi$). Moreover, each fault is equally likely to cause a failure.

The assumptions lead to the hazard rate $Z(\Delta t/ t_{i-1})$ after removal of the $(i – 1)^{st}$ fault being pro-portional to the number of faults remaining in the software $(N – M(t_{i-1}))$. Hence we have

$$Z(\Delta t/ t_{i-1}) = \varphi (N – M(t_{i-1})) \qquad (1)$$

The mean value function and the failure intensity functions for this model turns out to be

$$\mu(t) = N (1 – e^{-\varphi t}) \qquad (2)$$
$$\lambda(t) = N \varphi e^{-\varphi t} = \varphi (N – \mu(t)) \qquad (3)$$

Software reliability obtained from this model can then be expressed as

$$R(t_i) = e^{- \varphi (N – (i – 1)) t_i} \qquad (4)$$

The model requires the elapsed time between failures or actual failure times for estimating its parameters.

## 2.3 The Goel-Okumoto Model

The Goel-Okumoto (G-O) non-homogeneous Poisson process (NHPP) model has slightly different assumptions from the J-M model. The significant difference between the two is the assumption that the expected number of failures observed by time $t$ follows a Poisson distribution with a bounded and non-decreasing mean value function $\mu(t)$. The expected value of the total number of failures observed in infinite time is a finite value $N$.

The model also makes the following assumptions:
- The number of software failures that occur in $(t, t+\Delta t]$ is proportional to the expected number of undetected faults, $N - \mu(t)$.
- The number of failures detected in the inter-failure intervals $(0, t_1), (t_1, t_2), \ldots, (t_{n-1}, t_n)$ is not correlated.

- The per-fault hazard rate is time invariant and a constant ($\varphi$).
- The fault removal process when failures are detected is instantaneous and perfect.

The assumptions result in the following mean-value function (and corresponding failure intensity function) to be developed for the expected number of failures observed by time *t*.

$$\mu(t) = N(1 - e^{-\varphi t}) \qquad (5)$$
$$\lambda(t) = N\varphi e^{-\varphi t} = \varphi(N - \mu(t)) \qquad (6)$$

It turns out that for a Poisson process, the failure intensity equals the hazard rate and therefore we have the failure intensity for a fault equals the constant per-fault hazard rate $\varphi$. It is clearly seen from equations (2), (3) and (5), (6) that G-O model is mathematically equivalent the J-M model. The primary difference lies in the assumptions and the interpretation of what *N* is. In the J-M model, *N* is known and fixed, whereas in the G-O model, *N* itself is an expectation.

The model requires failure counts in the testing intervals and completion time for each test period for parameter estimation.

Several extrapolations of the G-O NHPP model exist (as do variants of the J-M model) incorporating different assumptions. The notable ones are those that consider different distributions for the times between failures and variable per fault hazard rate *e.g.* Yamada's S-shaped NHPP SRGM, or those that consider imperfect debugging [Pham00].

## 2.4 Musa's basic execution time model

Musa's basic execution time model is the first attempt at incorporating a time variant test effort into reflecting software reliability growth as testing is performed. This model is also an NHPP model and assumes that the number of failures observed by a time $\tau$ is finite and follows a Poisson distribution. A salient feature of this model is that $\tau$ represents the "execution time" in terms of actual CPU time utilized for the test that caused the failure.

Some of the important assumptions that the model makes are:
- The execution time between failures are exponentially distributed
- The per fault hazard rate is constant ($\varphi$)

The model requires that actual times of the failure be recorded as these times are used in the calculation of model parameters. The failure intensity obtained by application of the model to the failure data is

$$\lambda(t) = NB\varphi\, e^{-B\varphi t} \qquad (7)$$

Where *N* has the same interpretation as in the G-O model. *B* represents the fault reduction factor, which is a constant relating the rate of fault correction to the hazard rate. Equation (7) can also be put into the form

$$\lambda(t) = \varphi(N - \mu(t)) = Kf(N - \mu(t)) \qquad (8)$$

Where $\varphi$ is formulated as the product of the linear execution frequency *f*, and the fault exposure ratio *K*, which can be interpreted as the average number of failures occurring per fault remaining in the code during one linear execution of the program. The parameter is also assumed to be constant over time [Grottke01]. Equation (8) suggests that Musa's model is mathematically the same as the G-O and the J-M models. Indeed, this is the case, with the primary difference being the formulation of the per-fault hazard rate $\varphi$.

An interesting extension of this model is the logarithmic Poisson execution time model, where the expected number of failures is a Poisson random variable and a logarithmic function of the CPU time $\tau$, and a factor determining decay in failure intensity.

## 2.4   The enhanced NHPP model

The enhanced NHPP (ENHPP) model [Gokhale96] is a unifying framework for finite failure NHPP models *i.e.* other NHPP models with bounded mean-value functions are special cases of the ENHPP model. The model explicitly incorporates time-varying test coverage and imperfect fault detection in its analytical formulation.

Test coverage in this model is defined as the ratio of the number of potential fault sites sensitized by a test to the total number of potential fault sites. Potential fault sites refer to "the program entities representing either structural or functional program elements whose sensitization is deemed essential towards establishing the operational integrity of the software product" [Gokhale96].

The model makes the following assumptions:
- Faults are uniformly distributed over all potential fault sites.
- The probability of detecting a fault when a fault site is sensitized at time *t* is $c_d(t) = K$, (a constant), the fault detection coverage.
- Faults are fixed perfectly.

The mean value function for this model is developed as

$$\underline{\mu}(t) = c(t)\mathbf{N} \qquad (9)$$

Where $c(t)$ is the time variant test coverage function and $\mathbf{N}$ is number of faults expected to have been exposed at full coverage. This is distinguished from $N$, which is the expected number of faults to be detected after infinite testing time, perfect test and fault detection coverage. The failure intensity for this model then becomes

$$\lambda(t) = z(t)(\mathbf{N} - \mu(t)) \qquad (10)$$

Where $z(t) = c'(t).(1 - c(t))^{-1}$ is the time variant per-fault hazard rate. The model allows the scenario of defective coverage to be incorporated in the reliability estimation. Different coverage function distributions result in the variations of the NHPP models *i.e.* the G-O model, the Yamada S-shaped model, etc. Reliability as obtained from this model is expressed as

$$R(t/s) = e^{-NK(c(s+t) - c(s))} \qquad (11)$$

Where *s* is the time of last failure and *t* is the time measured from last failure. Grottke [Grottke01] observes correctly, that the main merit of this model is to serve as a unifying framework for NHPP models. Further, the dependence of the per-fault hazard rate solely on time-variant test coverage neglects other influencing factors such as the fact that full test coverage may not be successful in detecting all the faults and that failures may still occur without any gain in test coverage.

## 2.5 Littlewood – Verrall bayesian model

The models presented in the previous sections all assume that failure data is available. They also apply classical statistical techniques like maximum likelihood estimation (MLE) where model parameters are fixed but unknown and are estimated from the available data. The drawback of such an approach is that model parameters cannot be estimated when failure data is unavailable. Even when few data are available, MLE techniques are not trustworthy since they can result in unstable or incorrect estimations.

The bayesian SRGM considers reliability growth in the context of both the number of faults that have been detected and the failure-free operation. Further, in the absence of failure data, bayesian models consider that the model parameters have a prior distribution, which reflects judgement on the unknown data based on history *e.g.* a prior version and perhaps expert opinion about the software.

The Littlewood – Verrall model is one example of a bayesian SRGM that assumes that times between failures are independent exponential random variables with a parameter $\xi_i$, $i = 1,2, \ldots, n$ which itself has parameters $\psi(i)$ and $\alpha$ reflecting programmer quality and task difficulty) having a prior gamma distribution. The failure intensity as obtained from the model using a linear form for the $\psi(i)$ function is

$$\lambda(t) = (\alpha - 1)(N^2 + 2B\varphi(\alpha - 1))^{-\frac{1}{2}} \qquad (12)$$

Where $B$ represents the fault reduction factor, as in Musa's basic execution time model. This model requires tune between failure occurrences to obtain the posterior distribution from the prior distribution.

To conclude this section, almost all of the fundamental (non-bayesian) black-box models and their variations can be generalized to the model form of Equation (3) where failure intensity is proportional to the number of remaining faults in the software. [Lyu96] and [Pham00] provide a fairly comprehensive summary as well as a classification scheme for many of the commonly used SRGMs. A unifying view on SRGMs can also be found in [Grottke01].

## 3 White box software reliability models

White box software reliability models consider the internal structure of the software in the reliability estimation as opposed to black box models which only model the interactions of software with the system within which it operates. The contention is that black box models are inadequate to be applied to software systems in the context of component-based software, increasing reuse of components and complex interactions between these components in a large software system. Furthermore, proponents of white box models advocate that reliability models that consider component reliabilities, in the computation of overall software reliability, would give more realistic estimates.

The motivation to develop the so-called "architecture" based models includes development of techniques to analyze performability of software built from reused and commercial off-the shelf (COTS) components, performing sensitivity analyses *i.e.* studying the variation of appli-cation reliability with variation in component and interface reliability, and for the identification of critical components and interfaces [Gokhale98, Trivedi01].

In these white box models, components and modules are identified, with the assumption that modules are, or can be, designed, implemented and tested independently. The architecture of the software is then identified, not in the sense of the traditional software engineering architecture but rather in the sense of interactions between components. The interactions are defined as control transfers, essentially implying that the architecture is a control-flow graph where the nodes of the graph represent modules and its transitions represent transfer of control between the modules. The failure behavior for these modules (and the associated interfaces) is then specified in terms of failure rates or reliabilities (which are assumed to be known or are computed separately from SRGMs). The failure behavior is then combined with the architecture to estimate overall software reliability as a function of component reliabilities. The way in which the failure behavior is combined with the architecture suggests that three generic classes of white box software reliability models exist: path based models, state based models and additive models.

In the subsequent portions of this section, I describe Krishnamurthy and Mathur's path based model, Gokhale *et al.*'s state based model and briefly explain the underlying scheme in an additive model.

## 3.1 Krishnamurthy and Mathur's path based model [Mathur97]

The primary assumption in this model is that component reliabilities are known. The model averages path reliability estimates over all the test cases run on a particular software to estimate software reliability. Path reliability is computed from the sequence of components across a path that is followed when a particular test from the set of test cases is executed. The "architecture" of the software is essentially the sequence of components along different paths obtained from execution traces collected from testing or simulation of the software.

If each component has reliability $R_i$ then assuming that each component fails independently, the path reliability $R_p$ of a trace $M(P, t)$ for a program $P$ containing a sequence of components $m$ executed against a test case $t$ belonging to a test suite $T$ is

$$R_p = \prod_{m \in M(P,t)} R_i \qquad (13)$$

The overall system reliability $R_{sys}$ is the average of all path reliabilities over the test suite $T$ is

$$R_{sys} = \sum R_p \Big/ |T| \qquad (14)$$

Intra-component dependencies and the effect of a large number of loops introduce the possibility that multiple occurrences of a component along a trace path are not independent. This is modeled by considering that multiple occurrences of a component along the same path is equivalent to having $k > 0$ occurrences where $k$ represents the degree of independence. The larger the value of $k$ the lower is the estimate of path and consequently system reliability.

Other path based models follow similar approaches and system reliability is computed in general by considering possible execution paths of a program either by testing, experimentally or algorithmically.

## 3.2 Gokhale *et al.*'s state based model [Gokhale96]

The particular characteristic of this model is that it either assumes that component reliabilities are available or it determines component reliabilities using the ENHPP SRGM. Incidentally, the developers of this model and the ENHPP model are the same. Another assumption is that the application for which reliability is to be predicted is a terminating application. Gokhale *et al.* observe that if the control transfers between modules are assumed to be a markov process then the control-flow graph, describing the architecture, of the software can be directly mapped into a discrete-time or continuous-time markov chain, with a one-one correspondence between the architecture and the markov chain.

The transitions of the markov chain represent the transition probabilities between modules ($p_{ij}$) and the expected time spent in a module *i* per visit $t_i$ is computed as a product of the expected execution time of each block and the number of blocks in the module.

Component reliabilities are computed from the ENHPP model as

$$R_i = e^{-\int_0^{V_i T_i} \lambda_i(t)dt} \quad (15)$$

Where $V_i$ is the expected number of visits to module *I* and $\lambda_i(t)$ is the time dependent failure intensity and $V_i T_i$ is the total expected time spent in a module. The reliability of the overall system is then computed as

$$R_{sys} = \prod_i^n R_i \quad (16)$$

Yet another category of white box models contains models known as additive models. These focus on computing overall system reliability using component failure data and do not attempt to develop an architecture for the system. The overall system failure intensity is determined as the sum of component failure intensities assuming that component reliability can be estimated using the NHPP class of software reliability growth models.

There are several issues with regard to white box models that pose interesting research questions. One of the more important issues is discussed in the concluding section of this report. A more comprehensive summary of issues regarding architecture based models and several other white box models are fairly well documented in [Trivedi01].

## 4 Relating residual fault content to a worst case reliability estimate

In this section, I briefly describe a novel theory proposed in [Bishop96] and [Bishop02] that tries to relate fault content in software to worst-case failure rate and consequently a worst case bound on reliability. The proposed theory is applicable to black box SRGMs and makes the assumptions that:
- Removing a fault does not affect the failure rates of the remaining faults
- The distribution of inputs to the software is stable which implies that the failure rates of faults $\lambda_1, \lambda_2, \ldots, \lambda_n$ does not change with time
- The fault detection and correction process is perfect.

Bishop *et al*. observe that these assumptions are conservative but are the same as most other SRGMs make. The perceived (but unknown) failure rate ($\lambda_i$) is a function of the probability per unit time $P(j)$ of activating fault $i$ in the input space $D$ and is given by

$$\lambda_i = \sum_{j \in D} P(j) \qquad (17)$$

The assumptions that are made yield that a fault $i$ with a perceived failure rate $\lambda_i$ can survive a usage time $t$ with a probability $\exp(-\lambda_i t)$. The average failure intensity is therefore

$$\lambda'_i | t = \lambda_i \exp(-\lambda_i t) \qquad (18)$$

Whose maximum value is obtained at $\lambda_i = 1/t$ which in turn yields that the maximum failure intensity contribution of any fault after an operating time $t$ is

$$\lambda'_i | t \leq 1/et \qquad (19)$$

Where $e$ is the exponential constant. Equation (19) indicates that the result of maximum failure rate is independent of the actual failure rate for a fault. It then follows that for N faults in a program, the worst-case failure rate is simply

$$\lambda'_t \leq N/et \qquad (20)$$

Where $\lambda'_t = \Sigma \lambda'_i | t$. The worst case bound on reliability is then given as the probability $R(t/T)$ of operating for time $t$ without failure after prior usage $T$. This is evaluated as

$$R(t/T) = (1 - e^{-\lambda T}) + e^{-\lambda T} \cdot e^{-\lambda t} \qquad (20)$$

Where $\lambda = \lambda'_t$. The minimum value of $R(t/T)$ occurs at $\lambda = \ln(1+ t/T)t^{-1}$ which gives a lower bound of $t/(e(t + T)$ for $t << T$ and a lower bound of $T/t$ for $t >> T$. The notion of fractional fault is also introduced in this work where a fractional fault represents the possibility of certain implementations being defect free while others have defects. If imperfect diagnosis is considered then $d$ represents the number of failures that must occur before a failure is corrected. The theory then modifies the worst-case failure rate to $Nd/et$.

A limitation of the model is the assumption that the estimate of number of defects prior to testing is fixed. The predicted worst case bound is invalid if the detected number of defects is greater than the estimated value. The bounds that are predicted by this model however are pessimistic.

## 5  Observations and concluding remarks

With respect to using SRGMs for reliability estimation, consideration of the model assumptions is important before an SRGM is applied to failure data to ensure consistency between the model assumptions and corresponding data. For example, if a Weibull or a Gamma distribution fits the recorded failure times well, predictions obtained from a model that assumes a similar failure time distribution is more likely to be closer to actual values than a prediction from a model that assumes an exponential failure time distribution.

Further, quoting an observation made by Brocklehurst and Littlewood in [Lyu96], "There is no universally acceptable model that can be trusted to give accurate results in all circumstances; users should not trust claims to the contrary. Worse, we cannot identify a priori for a particular data source the model of models, if any, that will give accurate results; we simply do not understand which factors influence model accuracy". It is often the case that a group of growth models having similar assumptions disagree in their predictions for the same set of failure data and it is also the case that all the models make the same wrong prediction [Lyu96]. In such a scenario, the predictions from the models are moot and might only be best used for current reliability estimation rather than for prediction.

With regard to white box models, most models make the assumption that component reliabilities are available and ignore the issue of how they can be determined. This is still an open research issue. With scarcity of failure data in components, it is not always possible to employ SRGMs to estimate component reliabilities such as in *Gokhale et al.*'s state based model. Moreover, the assumption of independence between failures in components can be violated during unit testing, which implies that a reliability growth model can no longer be used to determine component reliabilities [Trivedi00]. Inter-component dependence is assumed to be nonexistent in architecture based models, which does not seem to be a very realistic assumption. The problem arises when an interface causes error propagation between two components and causes failures in both components. This invalidates the assumption of independence in component and interface failures and the models are no longer applicable.

The merit of architecture based models, especially of state based models, is primarily that the framework for reliability prediction can also be used for performance analysis, as well as for sensitivity analyses and in the identification of critical components.

Finally, most models rely on the existence of failure data with the exception of bayesian growth models that assume a prior distribution for the SRGM parameters. However, these models suffer from their inapplicability if software reliability is a function of the reliabilities of its components and interfaces. This appears to be the case with the increasing use of COTS in building software. Prediction of reliability at the testing stage allows for little feedback to the design process since testing is too far down the software engineering cycle.

In my view, a unifying framework that utilizes software metrics early during the software engineering cycle, failure data, when available, process metrics and process history to iteratively estimate or predict reliability would be of value in the sense of early validation of reliability requirements, for design tradeoffs and for evaluating software architectures. Further, no framework exists, yet, that produces a reasonable prediction of software reliability when data is lax and refines the prediction when data does become available. These are areas that merit further research.

# References

[Bishop96]   P.G. Bishop, R. Bloomfield, "A conservative theory for long term reliability growth prediction", *IEEE Transactions on Reliability, Vol. 45, No.4*, pp. 550-560, Dec. 1996.

[Bishop02]   P.G. Bishop, R. Bloomfield, "Worst case reliability prediction on a prior estimate

of residual defects", *Proceedings of the 13th IEEE International Symposium on Software Reliability Engineering (ISSRE-2002)*, pp. 295 – 303, Nov. 2002

[Gokhale96]  S. Gokhale, T. Philip, P. Marinos, K. Trivedi, "Unification of finite-failure non-homogenous Poisson process models through test coverage", *Proceedings of the 7th IEEE International Symposium on Software Reliability Engineering (ISSRE-96)*, Nov. 1996.

[Gokhale98]  S. Gokhale, W.E. Hong, K. Trivedi, J.R. Horgan, "An analytical approach to architecture based software reliability prediction", *Proceedings of the 3rd IEEE International Computer Performance and Dependability Symposium, (IPDS-98)*, pp.13-22, 1998.

[Grottke01]  M. Grottke, "Software reliability model study", *PETS Project Technical Report A.2*, University of Erlangen-Nuremberg, 2001.

[Lyu96]  M. R. Lyu (Ed.), *Handbook of Software Reliability Engineering*, IEEE Computer Society Press, 1996.

[Mathur97]  S. Krishnamurthy, A.P. Mathur, "On the estimation of reliability of a software system using reliability of its components", *Proceedings of the 8th IEEE International Symposium on Software Reliability Engineering (ISSRE'97)*, pp.146 – 155, Nov. 1997.

[Pham00]  H. Pham, *Software Reliability*, Springer-Verlag, 2000.

[Trivedi00]  K. Popstajanova and K. Trivedi, "Failure correlation in software reliability models", *IEEE Transactions on Reliability, Vol. 49, No. 1*, March 2000.

[Trivedi01]  K. Popstajanova and K. Trivedi, "Architecture based approach to reliability assessment of software systems", *Performance Evaluation, Vol. 45, No.2*, June 2001.